# Random Forest Stratified K-Fold Cross Validation on SYN DoS Attack SD-IoV

1st M. Arif Hakimi Zamrai
Faculty of Electrical Engineering
Universiti Teknologi Malaysia
81310 Skudai, Johor
mahakimi8@graduate.utm.my

2nd Kamaludin Mohd Yusof
Faculty of Electrical Engineering
Universiti Teknologi Malaysia
81310 Skudai, Johor
kamalmy@utm.my

3rd Muhammad Afizi Azizan
Faculty of Electrical Engineering
Universiti Teknologi Malaysia
81310 Skudai, Johor
muhammadafizi@graduate.utm.my

*Abstract—* **In response to the prevalent concern of TCP SYN flood attacks within the context of Software-Defined Internet of Vehicles (SD-IoV), this study addresses the significant challenge of network security in rapidly evolving vehicular communication systems. This research focuses on optimizing a Random Forest Classifier model to achieve maximum accuracy and minimal detection time, thereby enhancing vehicular network security. The methodology involves preprocessing a dataset containing SYN attack instances, employing feature scaling and label encoding techniques, and applying Stratified K-Fold cross-validation to target key metrics such as accuracy, precision, recall, and F1-score. This research achieved an average value of 0.999998 for all metrics with a SYN DoS attack detection time of 0.24 seconds. Results show that the fine-tuned Random Forest model, configured with 20 estimators and a depth of 10, effectively differentiates between normal and malicious traffic with high accuracy and minimal detection time, which is crucial for SD-IoV networks. This approach marks a significant advancement and introduces a state-of-the-art algorithm in detecting SYN flood attacks, combining high accuracy with minimal detection time. It contributes to vehicular network security by providing a robust solution against TCP SYN flood attacks while maintaining network efficiency and reliability.**

I. INTRODUCTION

Vehicular communication networks face network security challenges, notably Denial of Service (DoS) attacks, with TCP SYN flood attacks as a significant threat [1], [2], [3], [4]. Among these, TCP SYN flood attacks have emerged as a significant threat, exploiting the TCP three-way handshake mechanism to overwhelm network resources [1], [2], [5], [6]. These attacks are especially concerning in vehicular networks where they can disrupt critical communication, leading to potential safety hazards.

This research study uses machine learning to improve TCP SYN attack detection in vehicular networks, addressing the inadequacies of traditional security measures [1], [2], [3]. The approach is rooted in the realization that traditional network security measures are often inadequate in the dynamic and complex environment of vehicular communication. Therefore, this research proposes a Fine-Tuned Random Forest Classifier-based model, tailored to accurately process, and analyze vehicular network traffic data with high accuracy and low detection time of SYN flood attacks.

This research preprocesses a labelled network traffic dataset, including SYN attacks, to train the model. This study employs a range of preprocessing techniques, including feature scaling and label encoding, to optimize the dataset for effective learning. The study employs Stratified K-Fold cross-validation for a thorough model evaluation, focusing on key metrics like accuracy, precision, recall, and F1-score. A significant aspect of this approach is the emphasis on reducing the model's prediction time, acknowledging the time-sensitive nature of vehicular communication networks.

Furthermore, this research delves into the practical aspects of deploying such a machine learning model in real-world scenarios. This paper discusses the challenges of integrating this model within existing vehicular network infrastructures and the potential impact on network performance. The goal is to provide a solution that not only enhances the security of vehicular networks against TCP SYN flood attacks but also maintains the efficiency and reliability of these critical systems.

*A. The Importance of TCP Protocol in SD-IoV*

In the domain of vehicular communication, Transmission Control Protocol (TCP) is one of the most important components in vehicular communication [7], [8]. This research acknowledges various studies that have utilized TCP as the primary communication protocol in vehicular networks. TCP's dynamic capability is critical in Software-Defined Internet of Vehicle (SD-IoV) applications, where real-time analytics and decisions are important [1], [2], [3] TCP's bi-directional communication is crucial for Vehicle-to-Infrastructure (V2I) applications' data exchange needs as it is mainly used in roadside units and traffic control centers [8].

Based on the current state usage of Vehicular Ad-hoc Networks (VANETs), many routing protocols are based on the TCP/IP suite [7]. This includes innovative implementations, such as in Adwitiya's study [5], which incorporates PUSH flags in the TCP header to prioritize certain packets [5] to prioritize which packet get send first. Contrasting with User Datagram Protocol (UDP), TCP's reliability is further accentuated by its ability to guarantee packet delivery through acknowledgment messages.

IoV characterized by their dynamic nature, high mobility, and random network topology, present unique challenges for TCP. These include frequent connection failures, scalability issues, multi-hop data transfer complexities, and data loss concerns. The robustness of TCP in VANETs is also proved in its resilience against attacks that disrupt data transmission. High loss rates, which significantly impede TCP transmissions, lead to increased Retransmission Timeout (RTO) values, highlighting the protocol's sensitivity to network conditions [1].

Nonetheless, TCP's dynamic capabilities remain integral to vehicular communication, particularly for real-time data transfer in smart city contexts [2], [5], [9], [10], [11]. In smart city networks, communication between vehicles and RSUs is important for disseminating critical information about road conditions, accidents, and traffic flow. RSUs also play a crucial role in improving overall transportation and parking systems. In such scenarios, TCP's role in facilitating reliable and ordered data transmission becomes indispensable [2], [5], [6], [9], [10]. TCP's application in SD-IoV extends beyond traditional data transmission to more complex requirements like system and firmware updates in vehicles [12]. This research underscores the protocol's adaptability and resilience, making it a cornerstone in developing efficient, reliable, and secure vehicular communication systems in the context of SD-IoV.

*B. TCP Vulnerabilities to DoS Attack in SD-IoV*

In SD-IoV, TCP plays a critical role in reliable data communication. TCP's susceptibility to SYN DoS attacks presents significant challenges. This vulnerability is particularly concerning within the Software-Defined Networking (SDN) framework, which is increasingly being adopted in IoV for its flexibility and efficient management of network resources. TCP ensures the delivery of packets through acknowledgment messages but is exploited in SYN flood attacks. The attacker exploits the TCP handshake mechanism by sending numerous SYN requests without completing the connection with ACK responses. This leads to a flood of half-open connections, draining the server's resources and impairing its ability to handle legitimate traffic.

The implications of SYN DoS attacks in an SD-IoV context are profound in [2], [3], [8], [11]:

**Resource Strain on SDN Controllers:** SYN attacks overload SDN controllers in IoV, impairing network traffic management and risking system failure.

**Impact on Dynamic Network Topology:** These attacks exacerbate IoV dynamic topology, challenging SDN controllers' stability and efficiency.

**Degraded Network Performance:** SYN flooding consumes bandwidth and processing power, leading to latency, packet loss, and hampered IoV applications.

**Security Risks:** The centralized control of SDN in IoV becomes a target, with SYN attacks threatening network security and functionality.

**Scalability Concerns:** As IoV networks grow, SYN flooding poses increasing scalability challenges, emphasizing the need for advanced security measures.

## II. RELATED WORKS

DoS attacks are destructive forms of cyberattack that have been extensively studied in various contexts, particularly in SDN and traditional networks. These attacks typically involve numerous compromised devices, such as botnets or worms, to overwhelm a target network or service. Santhosh et al. propose a Protocol Dependent Detection and Classification System for detecting SYN Flood attacks. This method identifies abnormal patterns in power usage and request frequency, indicative of such attacks. By monitoring these metrics in real-time, the system can quickly detect and drop connections to prevent the attack [13].

Jia Rong et al. suggest a method involving the calculation of self-information associated with normal and abnormal packet pairs in a time domain. This approach helps to determine the network's state, whether safe, normal, abnormal, or overloaded [14]. Sehrish et al. present an entropy-based statistical approach for detecting and mitigating TCP SYN flood Distributed Denial of Service (DDoS) attacks. This approach is reliable and lightweight, with minimal false-positive rates. It employs a three-phased detection scheme, including monitoring incoming SYN packets using entropy standard deviation and a weighted moving average [2]. Chun Hao et al. have developed a method that combines cuckoo hashing with innovative whitelist/blacklist approaches to detect and mitigate SYN flooding attacks in SDN. This method is designed to improve detection accuracy and reduce the traffic and register size required in the network [3].

In the context of Apache Spark, Morfino et al. integrate various supervised machine learning methods into the MLlib library for efficient SYN-DoS cyberattack identification. Their study highlights the effectiveness of algorithms like Random Forest (RF), Decision Tree (DT), and others, with RF showing a 100% accuracy rate and the shortest training time for DT [15]. Zhong Ling et al. proposed a Double Deep Q-Network (DDQN) based DDoS detection method for IoV, utilizing a DDQN algorithm and a Kalman filter. While the specific detection time is not numerically stated, the study emphasizes the reduction of time consumption and convergence time, suggesting an improvement in DDoS detection speed [16]. Finally, Husein et al. studied and compared the performance of SVM, K-Means, and Neural Network algorithms using Snort, an intrusion detection system. The study found SVM to be the most effective for real-time DoS and probe attack detection, achieving an accuracy of 97.5% [17].

## III. MATERIALS AND METHODS

*A. SYN DoS Attack in SD-IoV*

SYN sequence is crucial in establishing and maintaining reliable connections in TCP's three-way handshake mechanism, and it is essential for initiating a connection between a client and a server. The process starts with the client sending a SYN packet to a car or base station as a node to request a connection. The receiver then acknowledges this request by sending back a SYN-ACK (synchronize-acknowledgment) packet. To complete the handshake, the client responds with an ACK (acknowledgment) packet. This exchange is crucial as it establishes a connection and synchronizes sequence numbers between the client and server, ensuring that data packets are transmitted in an orderly and reliable fashion. This methodical exchange of SYN, SYN-ACK, and ACK packets is a cornerstone of TCP's design, providing a dependable means of data transfer across networks in SD-IoV. It ensures that each side of the connection is aware of the other's existence, readiness to communicate, and agreement on initial sequence numbers, which are vital for tracking the data packets in the communication session.

However, the impact of SYN flooding in vehicular communication is destructive. Richard et al. noted that SYN attacks are more destructive on TCP compared to the UDP protocol [1]. This is because TCP, unlike UDP, guarantees the delivery of packets by sending acknowledgment messages. SYN attacks against TCP transmissions lead to significant data transmission disruptions and high Retransmission Time Out (RTO) values, severely affecting network reliability [1]. The SYN flooding attack is recognized as one of the most aggressive network security attacks, abusing the TCP handshake process to rapidly overwhelm a victim's memory storage. In traditional networks, this has often been mitigated by deploying firewalls in front of critical servers [3]. The relevance of this threat is accentuated in SD-IoV where many current routing protocols are TCP/IP-based [18]. IoV characterized by their high mobility and random topology, face significant challenges with TCP. These challenges include frequent connection failures, scalability issues, multi-hop data transfer, and data loss, which are exacerbated by SYN flooding attacks [18].

*B. DoS Attack Detection*

*1) Datasets*

The CIC-DDoS2019 dataset by the Canadian Institute for Cybersecurity encompasses 80 million network flows from 15 DDoS scenarios, including both legitimate and attack traffic. Detailing 88 features like IP addresses and packet sizes, it serves to enhance DDoS detection and mitigation research [19].

For this study, 82 features focusing on SYN normal and attack traffic were extracted to train machine learning models. The dataset underwent preprocessing, including data type conversions and elimination of redundant data. While the nature of DDoS attacks in SD-IoV mirrors traditional attacks, a key difference is their initiation by fast-moving vehicles, creating a distinct mobile DDoS threat [16]. Thus, the CIC-DDoS2019 dataset is employed to simulate DoS attacks in IoV contexts.

*2) Fine-Tuning Random Forest Hyperparameters*

To find an effective algorithm for DoS attack detection in SD-IoV, various machine learning techniques were evaluated, leading to the selection of Random Forest (RF) as the most suitable choice. RF stands out as an ensemble learning algorithm, synthesizing multiple decision tree models to form a more robust predictive model [20]. The uniqueness of RF lies in its approach: each decision tree within the forest is independently constructed, drawing on different random samples and feature subsets. This method significantly reduces the risk of overfitting, which is a common issue with single decision trees. By aggregating multiple trees, RF enhances the model's generalization capabilities, making it more reliable and versatile, as illustrated in Figure 1.

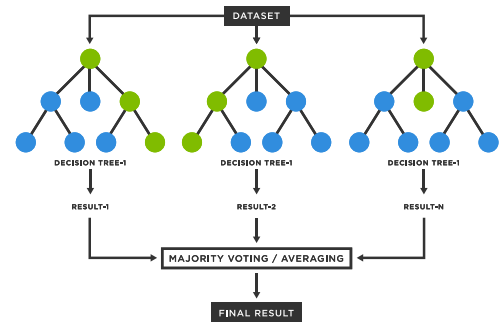

Fig. 1. Random Forest Algorithm Process

In Random Forest, the training of each decision tree involves a random subset of the original dataset, ensuring diversity in the learning process. The algorithm employs a voting or averaging mechanism to classify or regress, based on the collective output of these decision trees [21]. Key parameters of Random Forest include the sampling rate, the number of decision trees, and the feature count in each tree. Adjusting these parameters effectively balances the model's accuracy and computational efficiency. The choice of RF over other methods is influenced by several factors [21]:

**High Accuracy:** Random Forest typically achieves high accuracy in predictions, making it reliable for critical applications like DDoS attack detection.

**Feature Handling:** Random Forest can manage a large number of features and samples without compromising performance, a vital attribute given the complex nature of network data in DoS scenarios.

**Feature Importance Identification:** Random Forest is proficient at identifying which features are most significant in the prediction process, providing insights into the nature of the data and the attack patterns.

**Robustness to Missing Data:** Random Forest can handle datasets with missing values effectively, a common occurrence in real-world network traffic data.

**Diverse Applications:** Its versatility is evidenced by its successful application across various domains, including finance, healthcare, natural language processing, and computer vision.

In Algorithm 1, Random Forest hyperparameter tuning is a methodical process that begins with the CIC-DDoS19 datasets. These datasets undergo normalization through feature scaling using a standard scaler, essential for maintaining the model's integrity across varying feature scales. After scaling, Stratified K-Fold cross-validation is set up to ensure each fold reflects the overall class distribution, an important consideration given the characteristic class imbalance in DoS attack data. The number of estimators, the depth of the trees, and the feature selection method are the three critical hyperparameters the model explores. Estimators refer to the number of trees in the forest, depth specifies the maximum levels within each tree, and feature selection methods (sqrt, log2, or None) determine the subset of features considered at each split in a tree. An exhaustive search through combinations of these hyperparameters is conducted, with the goal of maximizing accuracy and minimizing prediction time. For instance, varying the number of estimators (with options such as 10, 20, 50, 100) affects the model's ability to learn from the data, potentially reducing bias but at the risk of increasing variance. Similarly, adjusting the tree depth (with options like 5, 10, 15, 20) balances the model's complexity against its generalization capabilities. The feature selection for each tree—whether it's the square root of the feature count (sqrt), the log2 of the feature count, or considering all features (None)—significantly impacts the diversity of the perspectives captured by the model.

Performance metrics such as Accuracy (Acc), Precision (Prec), Recall (Rec), F1 Score, ROC AUC, Prediction Time (T), and Confusion Matrix (Matrix) are recorded and analyzed to select the best-performing model configuration. This systematic evaluation allows for a nuanced understanding of the model's strengths and limitations, leading to the selection of the most effective combination of estimators, depth, and features for the final model. This method results in a model that is accurate and robust against overfitting while undergoing a comprehensive design and thorough validation process.

---

Algorithm 1: Fine-Tune Random Forest Algorithm

---

**Input:** CCIC-DDoS19 datasets, depth, estimator, feature options
**Output:** Fine-tune Random Forest model
**Procedure**: Fine-tuning parameters
  **Step 1:** Initialize [Acc, Prec, Rec, F1, ROC, Time, Matrix]
  **Step 2:** Feature Scaling scaler=StandardScaler()
  **Step 3:** Cross-Validation Setup
  Defined StratifiedKFold
  (n_splits=5, shuffle=True, random_state=42)
  **Step 4:** Model Evaluation Function `cross_val_model`
    For each train_index, test_index in skf.split(X, y)
      X_train_fold, X_test_fold = X[train_index], X[test_index]
      y_train_fold, y_test_fold = y[train_index], y[test_index]
      model.fit(X_train_fold, y_train_fold)
      predictions_fold = model.predict(X_test_fold)
      **Record**: [Acc, Prec, Rec, F1, ROC, Time, Matrix]
  **Step 5:** Optimize Number of Estimators
  estimator_options = [10,20,50,100]
  depth_options = [5, 10, 15, 20]
  feature_options = [sqrt, log2, None]
  For each combination in itertools.product(estimator_options, depth_options, feature_options)
    model = RandomForestClassifier(n_estimators, max_depth, max_features)
    [Acc, Prec, Rec, F1, ROC, Time, Matrix] =
    cross_val_model(model, X_scaled, y, skf)
    If accuracy > best_accuracy or (accuracy == best_accuracy and pred_time < best_pred_time)
      **Update**: best_accuracy, best_pred_time, best_estimators, best_depth, best_features
  **Step 6:** Final Model Selection
  Return the RandomForest model with the best combination
  of hyperparameters (best_estimators, best_depth, best_features).
  **Step 7:** Final Model Training
  Train the RandomForest model using the best
  hyperparameters on the entire training set.
  **Step 8:** Final Model Evaluation
  Evaluate the final model on the test set.
  **Output performance metrics:** accuracy, F1 score, precision, recall, and ROC AUC.

*C. Evaluating The Fine-Tune Random Forest Algorithm*

In this study, the fine-tuned Random Forest model's evaluation heavily relies on Stratified K-Fold cross-validation, which is particularly useful for imbalanced datasets like CCIC-DDoS19. This method divides data into several folds, each reflecting the dataset's overall class distribution, ensuring a balanced and thorough model assessment. Stratified K-Fold cross-validation lies in its ability to provide a more comprehensive and unbiased evaluation of the model [22], [23]. This rigorous evaluation process, along with the use of a confusion matrix to break down predictions into true positives, negatives, false positives, and negatives, underpins the model's practical efficacy. The study also focuses on metrics like accuracy, precision, recall, and F1 scores derived from the confusion matrix to comprehensively understand the model's classification capabilities. Equations (1), (2), and (3) delineate the definitions and formulas for these metrics. F1 score merges both precision and recall into a single metric. It is essentially a balanced mean of accuracy and recall, offering a comprehensive measure that accounts for both the precision of the classifier and its recall.

$$Precision = \frac{TP}{TP + FP} \quad (1)$$

$$Recall = \frac{TP}{TP + FN} \quad (2)$$

$$F1\ Score = 2 * \frac{Precision * Recall}{Precision + Recall} \quad (3)$$

IV. RESULTS AND DISCUSSION

The visualizations of the Random Forest model's performance metrics, such as accuracy, F1 score, recall, ROC AUC, and prediction time, offer insightful trends when viewed across different numbers of estimators and various feature selection methods used in constructing decision trees within the model. These methods, specifically 'sqrt', 'log2', and 'None', play important roles in determining how the model selects features at each split point during the learning process. The 'sqrt' option indicates that the model should randomly select the square root of the total number of features at each split. This method is based on the statistical principle that, given a high number of features, a subset of the square root of the total quantity often provides enough variability to create a strong learner yet is small enough to maintain computational efficiency. On the other hand, the 'log2' method suggests using the base-2 logarithm of the total number of features for selection at each split. This approach often leads to a greater bias but lower variance in the model, which can be beneficial in certain datasets where the number of features is very large and the model risks becoming too complex and overfitting the training data. 'None' as an option means no limit is set on the number of features; the model considers every feature at each split. This could lead to more complex models with higher variance but can be useful when the dataset is relatively small or when each feature significantly impacts the outcome. By examining the model's performance across these feature selection methods with varying numbers of estimators, one can deduce the most optimal configuration that balances bias and variance, thereby enhancing

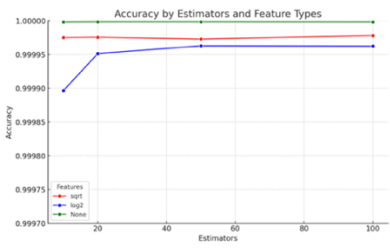
Fig. 2. Accuracy.

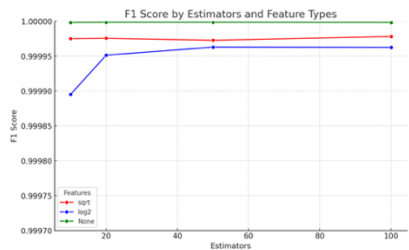
Fig. 3. F1 score.

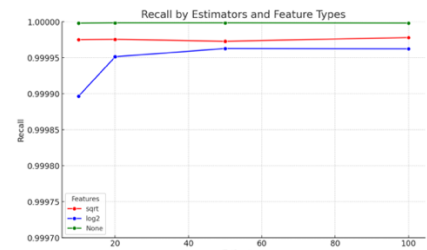
Fig. 4. Recall.

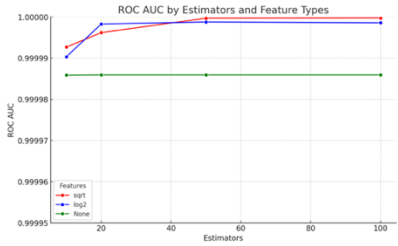
Fig. 5. ROC AUC.

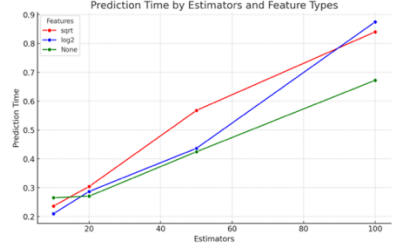
Fig. 6. Prediction Time.

the model's predictive power. Visualizing these trends is crucial, as it allows for a clear comparison of how each method affects the overall performance and speed of the model, guiding the fine-tuning process of the Random Forest algorithm in the context of DoS attack detection in vehicular networks.

In summary, as illustrated in Figure [2-6], key metrics such as accuracy, F1 score, recall, and ROC AUC generally improve when this study increases the number of estimators in a Random Forest model. The 'None' feature type often performs the best in these areas, particularly with a higher number of estimators, and it also strikes a good balance between effectiveness and quick prediction times. While the 'sqrt' feature type shows strong performance, but it require more time for predictions. The 'log2' feature type typically falls between 'sqrt' and 'None' in terms of performance and efficiency. Therefore, for a Random Forest model that prioritizes accuracy and efficiency, choosing the 'None' feature type with an increased number of estimators emerges as the optimal approach in this study.

### A. The Ideal Random Forest Hyperparameters

The optimal hyperparameters for the Random Forest model in this analysis are estimators 20 and depth 10, which exhibit a remarkable balance between high accuracy and low detection time. This is drawn from the detailed examination of various performance metrics using the fine-tuned Random Forest algorithm and the model's confusion matrix.

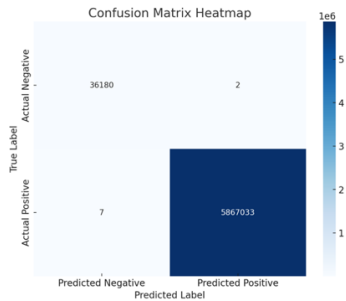
Fig. 7. Confusion Matrix Heatmap

The model achieves near-perfect accuracy, with a score of approximately 0.999998. Such high accuracy indicates the model's exceptional capability in correctly classifying both positive and negative instances. The F1 score, which is also around 0.999998, suggests a well-balanced model that excels in both precision (accuracy of positive predictions) and recall (ability to identify all positive instances). The precision score of nearly 0.999998 further emphasizes the model's accuracy in positive classifications, demonstrating that it rarely makes false positive errors. Similarly, the recall score mirrors this level of accuracy, indicating the model's effectiveness in capturing almost all positive cases.

Moreover, the model's ROC AUC value, approximately 0.99997, reflects its excellent ability to distinguish between the classes. This metric is crucial when distinguishing between positive and negative classifications is critical. In terms of computational efficiency, the model maintains a rapid prediction time of about 0.24 seconds. This speed is particularly impressive given the complexities usually associated with Random Forest models, especially those with a larger number of estimators or greater depth. The ability to deliver accurate results swiftly makes this configuration highly suitable for practical applications where accuracy and speed are valued. The model's effectiveness is further evidenced by its confusion matrix as in Figure 7, which reveals a very high number of true positives (5,867,033) and true negatives (36,180), coupled with an extremely low count of false positives (2) and false negatives (7). This matrix visually underscores the model's proficiency in making accurate classifications with minimal errors. The combination of estimators 20 and depth 10 for the Random Forest model stands out as the best configuration for this specific dataset and task. It provides an exemplary blend of high predictive performance across various metrics and efficient computational processing, making it a robust choice for real-world applications requiring precision and agility.

TABLE I. PERFORMANCE METRICS COMPARISON

| Author | Meth. | Acc. | Prec. | Rec. | F1-sc. | T(s) |
|---|---|---|---|---|---|---|
| Ma et al. [19] | RF | 0.9999 | 0.9999 | 0.9999 | 0.9999 | 0.4 |
| Li . et al. [16] | NN | 0.9740 | | | 0.9390 | |
| Husein et al. [17] | SVM | 0.9750 | | | | |
| This research method | FT-RF | 0.9999 | 0.9999 | 0.9999 | 0.9999 | 0.24 |

Based on Table 1, this research stands out for its comprehensive approach, offering significant improvements over other methods in several key aspects. Firstly, it **employs Stratified K-Fold Cross-Validation for more reliable and consistent performance metrics**, especially beneficial for datasets with imbalanced classes like 'CIC-DDoS2019'. This method **ensures a balanced and accurate evaluation by maintaining class distribution in each fold**. Secondly, the study adopts a holistic approach to hyperparameter tuning by evaluating **combinations of hyperparameters simultaneously**, unlike sequential methods used in [19]. This simultaneous tuning can uncover more effective model configurations by considering interactions between different parameters. Thirdly, this research study significantly **enhances detection time**, achieving a rapid response of only 0.24 seconds. This quick detection is critical in cybersecurity, particularly for SYN DoS attack detection in the dynamic SD-IoV environment. Lastly, the study **a comprehensive set of evaluation metrics**, including accuracy, F1 score, precision, recall, and ROC AUC. This approach provides a more detailed and nuanced assessment of the model's performance compared to studies with fewer or less diverse metrics. These combined factors position this research method as superior in evaluating and addressing cybersecurity threats.

## V. CONCLUSION

In conclusion, this research significantly advances SYN DoS attack detection in SD-IoV networks with its fine-tuned Random Forest model. This method excels with near-perfect metrics in accuracy, precision, recall, F1-score, and an impressive ROC AUC, highlighting its ability to accurately differentiate between normal and malicious traffic. The model's rapid prediction time, around 0.24 seconds, enhances its suitability for real-time applications. Utilizing Stratified K-Fold cross-validation ensures robust and reliable performance evaluation, especially for imbalanced datasets. This approach, combined with comprehensive hyperparameter tuning and a detailed evaluation using multiple metrics, establishes this model as a highly effective and efficient tool for SYN DoS attack detection. This research not only contributes significantly to the SD-IoV field but also sets a benchmark for future enhancements in network security, particularly in adapting to diverse and evolving cybersecurity threats in SD-IoV.